%% file: main.tex
\documentclass[12pt,a4paper]{paper}
\usepackage[left=2.3cm,top=2.0cm,right=2.3cm,bottom=2.0cm]{geometry}
\usepackage{graphicx}
\usepackage{comment}
\usepackage{lineno}

\usepackage[english]{babel}
\usepackage{graphicx}

\usepackage{amsfonts}

\usepackage{hyperref}
\graphicspath{{logos/}{figures/}}

\usepackage{cite} % Allows for ranges in citations
\usepackage{mciteplus}

\usepackage{color}
\usepackage[normalem]{ulem}
\begin{document}
%\maketitle

%%%%%%%%%%%%%%%%%%  Authors  %%%%%%%%%%%%%%%%%%
\begin{center}
\begin{Large}
{\bf Paarl Africa Underground Laboratory}
\end{Large}
\vspace{1. cm}

\noindent Robert~Adam$^{5,1}$, Claire~Antel$^{14}$, Munirat~Bashir$^{23}$, Driss~Benchekroun$^{18}$, Xavier~Bertou$^{20}$, Markus~Böttcher$^{8}$, Andy~Buffler$^{7}$, Andrew~Chen$^{4}$, Rouven~Essig$^{22}$, Jules~Gascon$^{12}$, Mohamed~Gouighri$^{19}$, Trevor~Hass$^1$, Gregory~Hillhouse$^6$, Abdeslam~Hoummada$^{18}$, Anslyn~John$^1$, Pete~Jones$^3$, Youssef~Khoulaki$^{18}$, Luca~Lavina$^{13}$, Lerothodi~Leeuw$^2$, Mantile~Lekala$^9$, Robert~Lindsay$^2$, Roy~Maartens$^2$, Yin-Zhe~Ma$^1$, Fairouz~Malek$^{{11},*}$, Peane~Maleka$^3$, Jacques~Marteau$^{12}$, Rachid~Mazini$^{21}$, Thebe~Medupe$^{8}$, Bruce~Mellado~Garcia$^4$, Marcello~Messina$^{15}$, Lumkile~Msebi$^2$, Chilufya~Mwewa$^{26}$, Zina~Ndabeni$^{3,7}$, Richard~Newman$^1$, George~O'Neill$^{16}$, Fabrice~Piquemal$^{10}$, Lydia~Roos$^{13}$, Daniel~Santos$^{11}$, Silvia~Scorza$^{11}$, Fedor~Simkovic$^{24}$, Ivan~Stekl$^{25}$, Yahya Tayalati$^{17}$, Smarajit~Triambak$^2$, Zeblon~Vilakazi$^4$, Shaun~Wyngaardt$^1$,  JJ~van~Zyl$^1$ \\
\vspace{1. cm}

%June 21$^{st}$, 2023 (Version submitted to arXiv)
\end{center}
\vspace{1. cm}

%%%%%%%%%%%%%%%%%%   Affiliations  %%%%%%%%%%%%%%%%%%
\noindent$^1$Stellenbosch University-South Africa; \\ 
$^2$University of the Western Cape-South Africa; \\
$^3$iThemba LABS-South Africa;\\
$^4$School of Physics, University of the Witwatersrand Johannesburg-South Africa;\\
$^5$Square Kilometre Array Observatory-South Africa;\\
$^6$Botswana International University of Science and Technology-Botswana;\\
$^{7}$University of Cape Town-South Africa;\\
$^{8}$North West University Potchefstroom-South Africa;\\
$^9$The University of South Africa;\\
$^{10}$LP2I, CNRS-IN2P3, Université Bordeaux-France;\\
$^{11}$LPSC, CNRS-IN2P3, Université Grenoble Alpes-France;\\
$^{12}$IP2I, CNRS-IN2P3, Université Claude Bernard Lyon-France;\\
$^{13}$LPNHE, CNRS-IN2P3, Sorbonne Université Paris-France;\\
$^{14}$ Université de Genève-Switzerland;\\
$^{15}$ LNGS, Gran-Sasso-Italy;\\
$^{16}$European Spallation Source ERIC, Lund,-Sweden; \\
$^{17}$Mohammed V university of Rabat-Morocco;\\
$^{18}$Hassan II university of Casablanca-Morocco;\\
$^{19}$Ibn Tofail University of Kenitra-Morocco; \\
$^{20}$Centro At\'omico Bariloche, CNEA/CONICET-Argentina;\\
$^{21}$Institute of Physics, Academia Sinica, Taipei-Taiwan;\\
$^{22}$Stony Brook University, USA;\\
$^{23}$Ibrahim Badamasi Babangida University-Nigeria;\\
$^{24}$Comenius University Bratislava-Slovakia;\\
$^{25}$IEAP CTU Prague-Czechia;\\
$^{26}$Brookhaven National Laboratory, USA.
\vspace{0.5 cm}

\noindent\small{*Contact editor:fmalek@in2p3.fr}

\clearpage
\hrule
\begin{abstract}
Establishing a deep underground physics laboratory to study, amongst others, double beta decay, geoneutrinos, reactor neutrinos and dark matter has been discussed for more than a decade within the austral African physicists’ community. PAUL, the Paarl Africa Underground Laboratory, is an initiative foreseeing an open international laboratory devoted to the development of competitive science in the austral region. It has the advantage that the location, the Huguenot tunnel, exists already and the geology and the environment of the site is appropriate for an experimental facility. The paper describes the PAUL initiative, presents the physics prospects and discusses the capacity for building the future experimental facility. 
\end{abstract}
\hrule
\vspace{0.5 cm}

%\tableofcontents
%\addtocontents{toc}{\protect\setcounter{tocdepth}{2}}
%\clearpage
%-------------------------------------------------------------------------------
% 1 Introduction
\input{introduction}
%-------------------------------------------------------------------------------
%\clearpage
%-------------------------------------------------------------------------------
% Research purpose
\input{purpose}
%-------------------------------------------------------------------------------
%\clearpage
%-------------------------------------------------------------------------------
% Landmark Physics/Flagship experiments   
\input{astrophysics}
%-------------------------------------------------------------------------------
%\clearpage
%-------------------------------------------------------------------------------
% Landmark Physics/Flagship experiments   
\input{facility}
%-------------------------------------------------------------------------------
% \clearpage
%-------------------------------------------------------------------------------
% Conclusion
\input{conclusion}
%-------------------------------------------------------------------------------
%\clearpage
%-------------------------------------------------------------------------------

\input{references}
\end{document}

%% file: introduction.tex
\section{Introduction}
\label{sec:intro}

Neutrinos and the search for dark matter have been big drivers in the field of experimental physics, sitting at the frontier between nuclear physics, particle physics, astroparticle physics and cosmology. These studies are only possible in underground laboratories (ULs) where the experiments are shielded from cosmic rays by at least $\sim$~1000~m of rock. These laboratories are located in mines and tunnels, and most of the operating laboratories are to be found in the northern hemisphere: Europe, USA, Russia, Canada and Japan. Indeed, Europe is an example of such an integration with a network of investigation programs, common laboratories and integrated systems for education and investigation, allowing knowledge transfer to its institution and the society.
\begin{figure}[bth]
\centering
\includegraphics[width=1.0\textwidth]{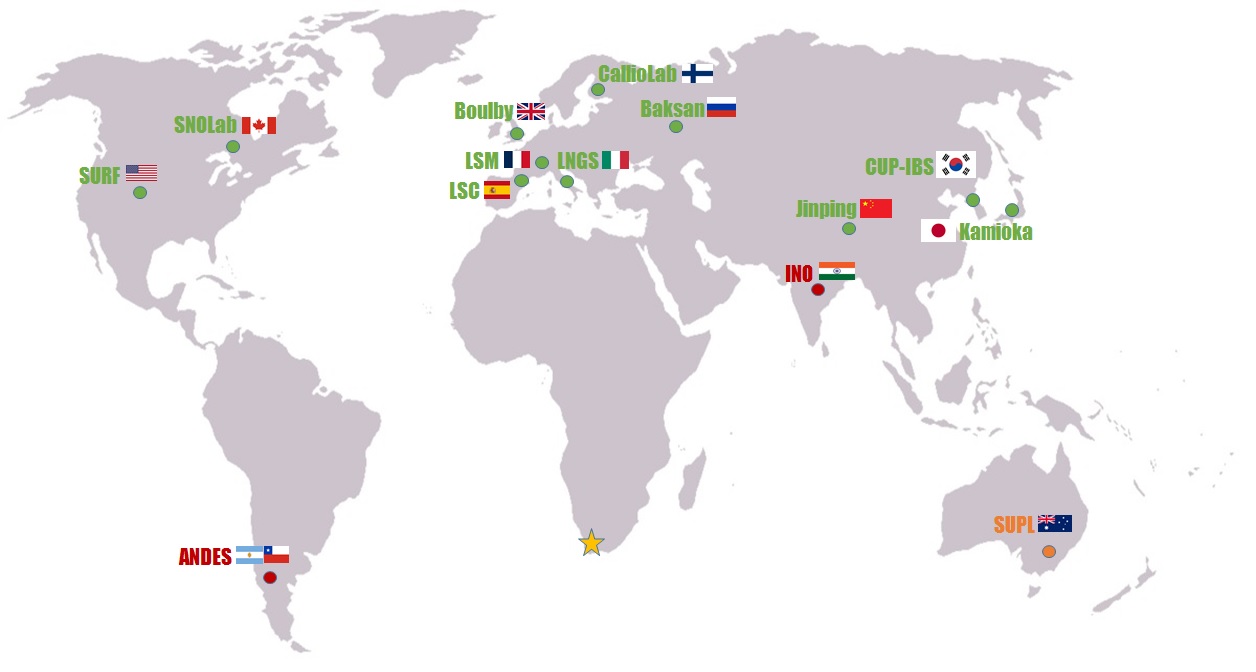}
\caption{\label{fig:ulabs} \small{Map of the existing or planned underground laboratories. Green dots: the operating facilities. Orange dot: under commissioning SUPL facility in Australia . Red dots: ANDES at the Argentina-Chile border and INO in India. Yellow star: the future Paarl Africa underground lab facility, in the Western Cape Province of South Africa}}
\end{figure}

Figure~\ref{fig:ulabs} shows the map of the existing or planned underground laboratories. At present, there are 11 ULs in operation (green dots), one is under commissioning (orange dot, SUPL in Ausralia), and two more (red dots) are planned: ANDES at the Argentina-Chile border and INO in India. ANDES (Ague Negra Deep Experiments Site)\cite{andessite} aims to create a 4000\,m$^2$ general purpose deep underground laboratory with 1750\,m of overburden in the Agua Negra tunnel between Argentina and Chile. INO (India-based Neutrino Observatory)\cite{inosite} is a planned laboratory in Pottipuram, India, with 1200\,m of overburden to host an iron calorimeter (ICAL) for neutrino studies. The yellow star is the foreseen location of the future Paarl Africa underground lab facility, in the Western Cape Province of South Africa, under the 1300~m Du Toitskloof mountain with $\sim$~800~m of rock overburden for the Huguenot tunnel itself. About 100 experiments are running or are under construction in the ULs currently in operation and roughly 6000 researchers are involved.

\begin{figure}[bth]
\centering
\includegraphics[width=0.9\textwidth]{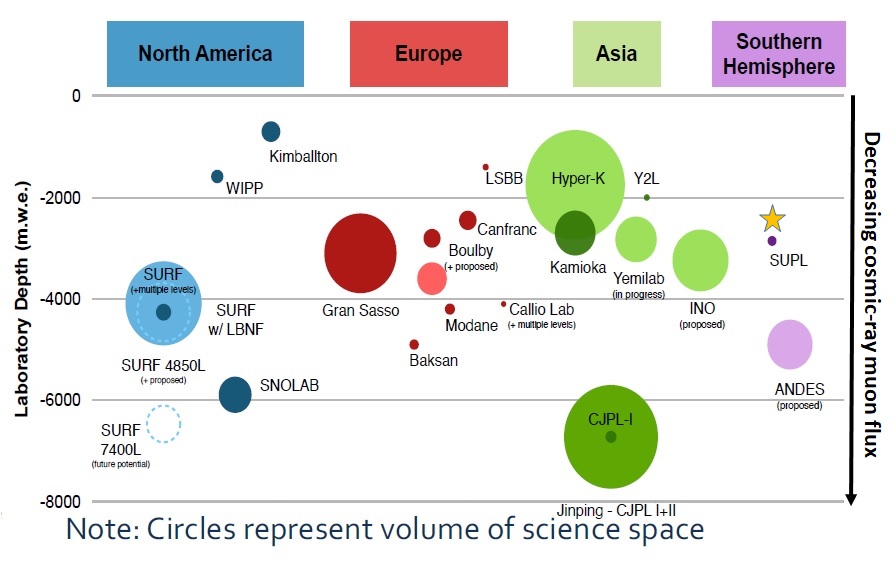}
\caption{\label{fig:flux} \small{Effective depth (in meter water equivalent\cite{mwe}) vs decreasing cosmic-ray muon flux~(cm$^2$~s$^{-1}$)\cite{Ianni}}. The yellow star indicate the corresponding numbers for the future Paarl Africa underground lab facility in the Western Cape Province of South Africa, under the 1300~m Du Toitskloof mountain with $\sim$~800~m of rock overburden for the Huguenot tunnel.}
\end{figure}

From the experimental point of view, the events that may lead to the identification of dark matter particles and rare electroweak decays have extremely low counting rates and would be observed at energies where an overwhelming background of radioactive decays and cosmic rays is present. Given the high energy and penetration of the cosmic radiation, the detectors need to be shielded by going underground. One can get a reduction of the cosmic radiation effects by orders of magnitude by digging labs some kilometers under the surface, or by building them under high geological features - see Figure~\ref{fig:flux} showing the laboratory depth (in meter water equivalent \footnote{The meter water equivalent (m.w.e.)\cite{mwe} is a standard measure of cosmic ray attenuation in underground laboratories. A laboratory at a depth of 1000 m.w.e is shielded from cosmic rays equivalently to a lab 1000 m below the surface of a body of water.}) as a function of the cosmic-ray muon flux for each continent. Circles represent the volume of science space.

The scientific work can expand to research projects in multiple fields including biology, geosciences, chemistry, mining technology and underground construction and architecture.

Discussions about an underground research facility in South Africa started in 2011. As one of the world’s largest producers of gold, South Africa has a number of the world’s deepest gold mines, including the recently-closed TauTona Gold Mine with a depth of 3900~m, and Mponeng deeper still. In 1965 the Nobel Prize laureate Frederich Reines along with South African Physicist, Friedel Sellschop detected the first atmospheric neutrino events in the East Rand mine near Johannesburg, South Africa (SA)~\cite{nobel}. Initial focus by the South African nuclear physics community was on establishing an underground facility in one of South Africa's deep gold mines. The alternative is to develop such an underground laboratory inside the Huguenot tunnel, which is conveniently located between the towns of Paarl and Worcerster in the Western Cape Province of South Africa, see Figure~\ref{fig:ulabs} and section~\ref{sec:paul}.

%% file: purpose.tex
\section{Research Purpose}
\label{sec:research}

The underground laboratory will provide a dynamic environment for advances in ultra-sensitive detectors and ultra-low radiation techniques and highly trained graduates ready to lead innovation in both the global search for \textbf{rare events  and cutting-edge technological development} to benefit South Africa industry. However, physics research is not the only field of interest. 

The need for very low radioactive material for dark matter and neutrino underground experiments gave birth to the study of new detectors able to measure \textbf{extremely low radiation levels}. These very sensitive detectors, able to detect levels of radiation a millionth of the natural radiation of the human body, have to be located deep underground to be shielded from cosmic radiation. The industry has shown interest in these techniques to select pure materials with almost no radioactive content. Researchers involved in this work can contribute to many needs in South Africa for accurate measurements, such as the detection of the radioactive gas radon that has been identified as a major radiation hazard in South African underground mines~\cite{radiation}.

The research of endolithic bacteria and technologies for bio-leaching are a subject of continuously increasing interest. New ideas in \textbf{underground biology}\cite{bio1,bio2,bio3} have also included initiatives related to \textbf{astrobiology}\cite{astrobio}, examining the impact of radiation (or the lack of it) to evolutionary processes or formation of bio-aerosols. 

The possibility to control the lighting conditions and other environmental parameters makes underground laboratories ideal places to experiment with \textbf{hydroponic farming}. They provide an \textbf{ideal sandbox to experiment with lighting}, determining the ability to use underground conditions as a working environment and even landscaping the tunnels as habitable environments, particularly with the very stable temperatures that occur underground.

The underground laboratory can also provide new insight on \textbf{environmental topics}. In glaciology, the study of ice samples from the Arctic, Antarctic etc. allows mapping of the evolution of climatic parameters and contamination both in space and over time for the last centuries. The measurement of $^{137}$Cs and $^{241}$Am is the only way to get a precise dating of ice samples and has to be done in underground laboratories\cite{ice}.

Underground laboratories are excellent sites to investigate new technologies for geothermal energy production or energy storage as well as the characterization of the geology of the earth\cite{neutrinogeo}. The Cape Supergroup, where the lab would sit, has been identified as a region of interest for geothermal research in particular.

%% file: astrophysics.tex
\section{Landmark Physics and Flagship astroparticle experiments}
\label{sec:physics}

\begin{figure}[bth]
\centering
\includegraphics[width=0.7\textwidth]{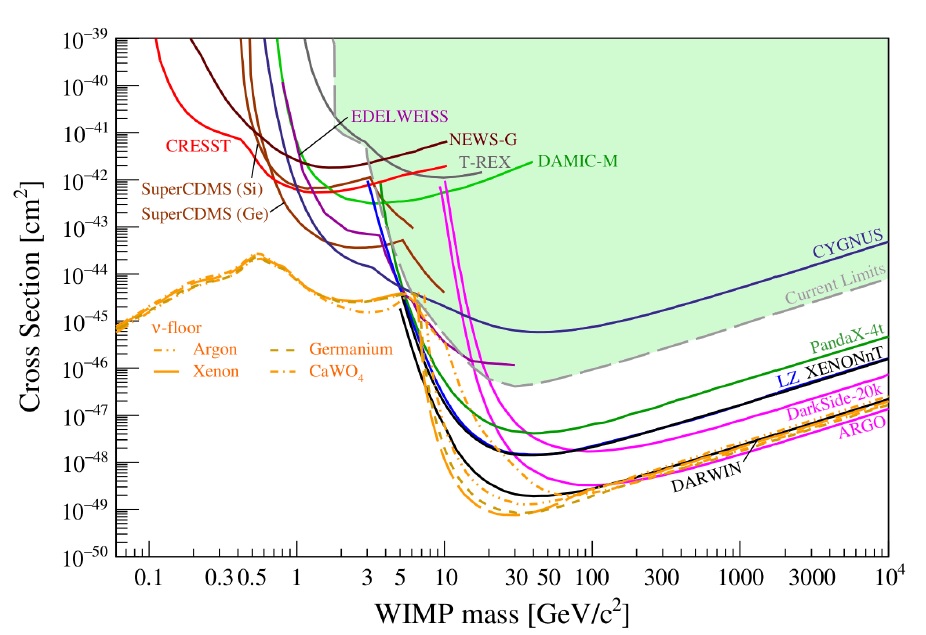}
\caption{\label{fig:wimps} \small{Current status of the Spin-Independent WIMP-nucleon interaction, one of the most common models used to look for WIMPs\cite{wimps}.}}
\end{figure}

Underground laboratories provide the low-background radioactive environment necessary for \textbf{astroparticle physics} to explore extremely rare phenomena. The underground location naturally guarantees high suppression of muons and cosmic-ray particles produced in the atmosphere and, consequently, of cosmogenic by-products. 

In the case of astroparticle physics, the leading fields are the direct search for dark matter and the study of neutrino properties. Figure~\ref{fig:wimps} shows the current landscape for direct dark matter searches~\cite{wimps}. The figure is scoping the WIMP (Weakly Interacting Massive Particle) model (one of the most promising models in which dark matter is typically assumed to have been produced thermally in the early universe) and parameters are constrained by the measured relic abundance of dark matter today. When looking for very low cross-sections, those detectors will soon start detecting neutrinos (from our Sun, from the atmosphere, from supernovae, etc…). This so-called neutrino floor, acting as background, will make the dark matter search more challenging, but it will also open new doors to neutrino physics studies. As can be seen from the figure, two regions of the parameter space still remain to be explored, which require different experimental efforts.

At high WIMP masses, only detectors using noble liquids (Xenon and Argon) can reach the required sensitivity. Besides the continuous reductions of the radiogenic background, the next generation experiments aim to build larger detectors to reach the neutrino floor, often surrounded by even more large shielding systems. This implies larger volumes and, as a consequence, wide footprints to host all subsystems. The timescale of next generation dark matter experiments foresees data taking ranging from 2030 (DARWIN~\cite{darwin}, seen as the natural step after the currently running experiments XENONnT~\cite{xenonnt} and LZ~\cite{lz}) and beyond (Argo, which is the next step of DarkSide-20k~\cite{darkside}, still under construction). While the underground site for those experiments is not yet defined, a novel underground site that does not surpass the existing ones in terms of depth can hardly be a good choice for them.

At small dark matter masses, however, there are many new opportunities to which a novel underground laboratory can contribute (for recent reviews with references see e.g.~\cite{Battaglieri:2017aum,Essig:2022dfa}).  Dark matter with mass below the proton (sub-GeV) typically lies in a dark sector, which does not interact directly with any of the Standard Model forces.  Instead, new particles (such as dark photons, scalars, or pseudo-scalars) can mediate interactions between the dark matter and the ordinary matter.  Dark matter particles can scatter off electrons, nuclei (via the Migdal effect), or collective modes in a detector target, leaving small signals consisting of ionization or phonons.  

There is nowadays a plethora of experimental techniques that are trying to gain sensitivity to such small signals. In this case, the challenge is to develop detectors with very low energy thresholds and excellent control over detector backgrounds, rather than to build large detectors that are highly demanding in terms of occupied volumes in an underground laboratory. In fact, the last few years has seen significant efforts in the direct-detection community on building small-scale experiments that can probe low-mass dark matter. 

A range of target materials, including liquid xenon, liquid argon, superfluid helium, silicon, germanium, gallium-arsenide, sapphire, diamonds, molecular liquids and gases, quantum dots, etc. are being used, or have been proposed, as target materials. 
For example, the  DarkSide~\cite{darkside} collaboration plans to have a "Low Mass" version of their technology (DarkSide-LM), tuned to have a radiogenic background optimised for low mass searches, using only one-ton of liquid Argon. So far, there is no established schedule on when and where this detector will be built. This could be a good opportunity for PAUL. 

Another technology that has recently come to the forefront is the Skipper-CCD~\cite{Tiffenberg:2017aac}, which has achieved single-electron sensitivity. Charge Coupled Devices (CCDs) are known to be very useful in digital cameras and widely used in research for it. Experiments like SENSEI~\cite{sensei} and DAMIC-M~\cite{damic} have already achieved several  results using small detectors~\cite{Crisler:2018gci,SENSEI:2019ibb,SENSEI:2020dpa,DAMIC-M:2023gxo}, and are now working to increase the target mass. Both collaborations are planning a next-generation detector called Oscura~\cite{oscura,Oscura:2022vmi}. 

Cryogenic detectors, operating at temperature well below 1 Kelvin, have also achieved a sensitivity at the single-electron level~\cite{HVeV,Edelweiss-DMe}. At the same time work is done to extend the particle-identification capabilities offered by this technology to effectively control the new backgrounds that appear as new energy domains are explored~\cite{excess}.

There are many other efforts and plans, and since the hunt for low-mass dark matter is relatively young, there is space for new experiments and new underground sites. The search for light dark matter particle is particularly attractive for these new technologies as significant advances can be achieved with kg-scale detector arrays.

One of the most interesting facts about having the possibility to perform an experiment of direct dark matter detection in an underground laboratory located in the Southern Hemisphere is to compare the eventual systematic errors or modulation with respect to the same detector in the Northern Hemisphere. Any systematic error or annual modulation correlated to a seasonal variation will have an opposed phase, giving the opportunity to discriminate them with respect to a dark matter signal. It also opens different regions of parameter space when searching for daily modulations~\cite{daily1,daily2,Avalos:2021fxm}. 

The other advantage to build an UL facility in South Africa is to combine the direct detection with indirect dark matter detection from radio astronomy surveys that South Africa is leading. As is well known, South Africa has been leading the world-astronomy collaboration ``Square Kilometre Array'' (SKA) mid-frequency arrays~\cite{SKA_link}, and has already built 64-dishes precursor ``MeerKAT'' telescopes in 2018~\cite{MeerKAT_link}. The MeerKAT telescopes are equipped with L-band ($856$-$1711$\,MHz) and UHF-band ($544$-$1087$\,MHz) receivers with average channel width $16\,$kHz. The dark matter (WIMP) annihilation into standard model particles (e.g. $b\bar{b}$, $\tau^{+}\tau^{-}$, $\mu^{+}\mu^{-}$) can eventually cascade to $e^{+}e^{-}$ pairs, which can lose their energies by inverse Compton-scattering and synchrotron radiation. This cascade process can generate fluxes in X-rays and radio wavelength respectively, with detailed variation determined by the dark matter mass and the astrophysical environment. In Ref.~\cite{Guo2023}, we have shown that with L-band receiver the radio antenna can place constraints on the WIMP cross-section in the mass range of $\sim 10$-$100$\,GeV. In addition, the annihilation and decay of WIMP during dark ages can leave imprints in the cosmic microwave background (CMB) radiation and alter the cosmic reionization proces~\cite{Cang2020}, which can be measured by high-sensitivity Hydrogen-Epoch Reionization Array (HERA~\cite{MeerKAT_link,reionization_web}) and CMB Stage-4 survey~\cite{cmbs4_web}. Both experiments have South Africa's deep participation and involvement. Therefore, the strong synergy between the astrophysical (indirect) probes and Paarl Africa Underground Laboratory (direct probe) can jointly measure and constrain dark matter effect, which may shed lights on new physics.

%% file: facility.tex
\section{The Paarl Africa Underground Laboratory and the experimental facility}
\label{sec:paul}

The development of the Huguenot tunnel as an underground low-level radiation facility holds a number of strategic advantages. Such a facility is located approximately 25~km from Stellenbosch University and 40 km from the iThemba LABS. It offers a unique possibility to build up a scientific complex suitable for the detection of events under ultra-low radiation exposure, as required for dark matter and neutrino studies, as well as for other areas of research. 
\begin{figure}[bth]
\centering
\includegraphics[width=0.7\textwidth]{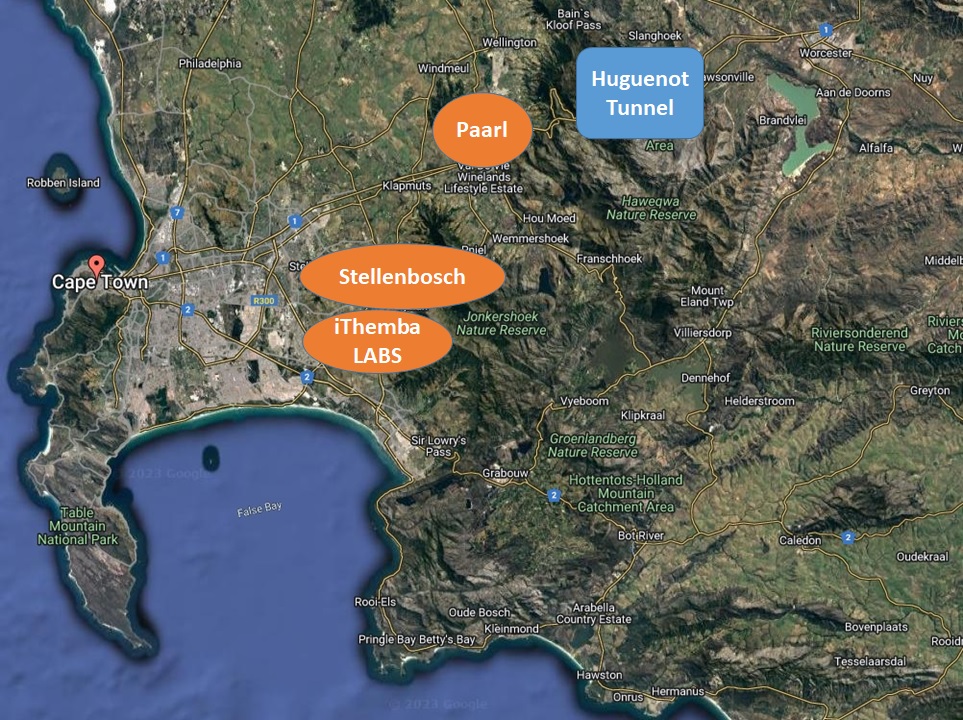}
\caption{\label{fig:CTMap} \small{Map of the Western Coast of Cape Town and the location of the Huguenot tunnel. Modified image extracted from Google Map.}}
\end{figure}

The tunnel provides quick and easy access to the local research communities. Research programs done at such a facility will also support postgraduate training programs at Stellenbosch University, the University of the Western Cape and the University of Cape Town. Furthermore, the research at the PAUL will support national and international research activities in astronomy, nuclear and particle physics, as well as many of the research topics already discussed in section~\ref{sec:research}.

\begin{figure}[bth]
\centering
\includegraphics[width=1.0\textwidth]{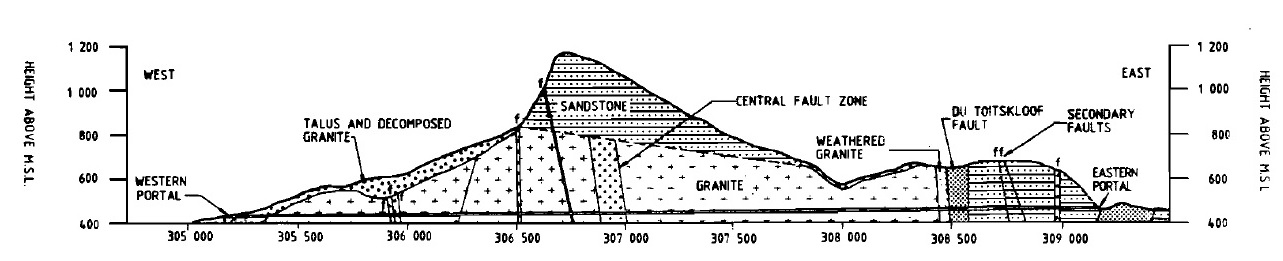}
\caption{\label{fig:geology} \small{Geological mapping and analysis of the rock of the tunnel~\cite{geo}.}}
\end{figure}

According to the geological survey, which was done during the excavation period (1973 - 1984), the composition of the mountain range consists mainly of quartzitic sandstone (also referred to as Table mountain sandstone), see Figure~\ref{fig:geology}. It is therefore expected that radon concentration levels within the tunnel should be low, and should therefore not have a considerable contribution to the background radiation signature. The height of the Du Toitskloof mountain range is about 1300~m, providing $\sim$~800~m of rock overburden for the Huguenot tunnel itself~\cite{geo}.

In 2013, the environmental radiation groups at Stellenbosch University and iThemba LABS performed a preliminary radon measurement~\cite{saul}.The concentrations measured at three sites confirm that the level of radon is well below any degree of consideration, with a mean level of radon no more than 64.9~Bqm$^{-3}$ (others showing mean levels as low as 45.4~Bqm$^{-3}$). In addition to the gamma ray spectra and radon monitoring, the background measurements were performed alongside biomonitors placed at strategic locations within the tunnel in order to measure the air pollution levels within the tunnel.

There is a real opportunity for a dedicated underground research facility to be established inside the Huguenot tunnel. The ideal would be to build the laboratory at the north of the service tunnel~\cite{huguenot} (to be built itself to the north of the current one, driving from Paarl to Rawsonville). The laboratory will consist of a surface facility, located near the road freeway Huguenot tunnel, and potentially extensive underground spaces and various connecting tunnels. The experimental hall will be covered by about 800~m of rock, under the Du Toitskloof Mountains,  protecting the experiments from cosmic rays. It will be easily accessible, with the ability to drive to it rather than using mine-shaft elevators, similarly to the LSM~\cite{lsm} or LNGS~\cite{lngs} facilities.

\begin{figure}[bth]
\centering
\includegraphics[width=0.9\textwidth]{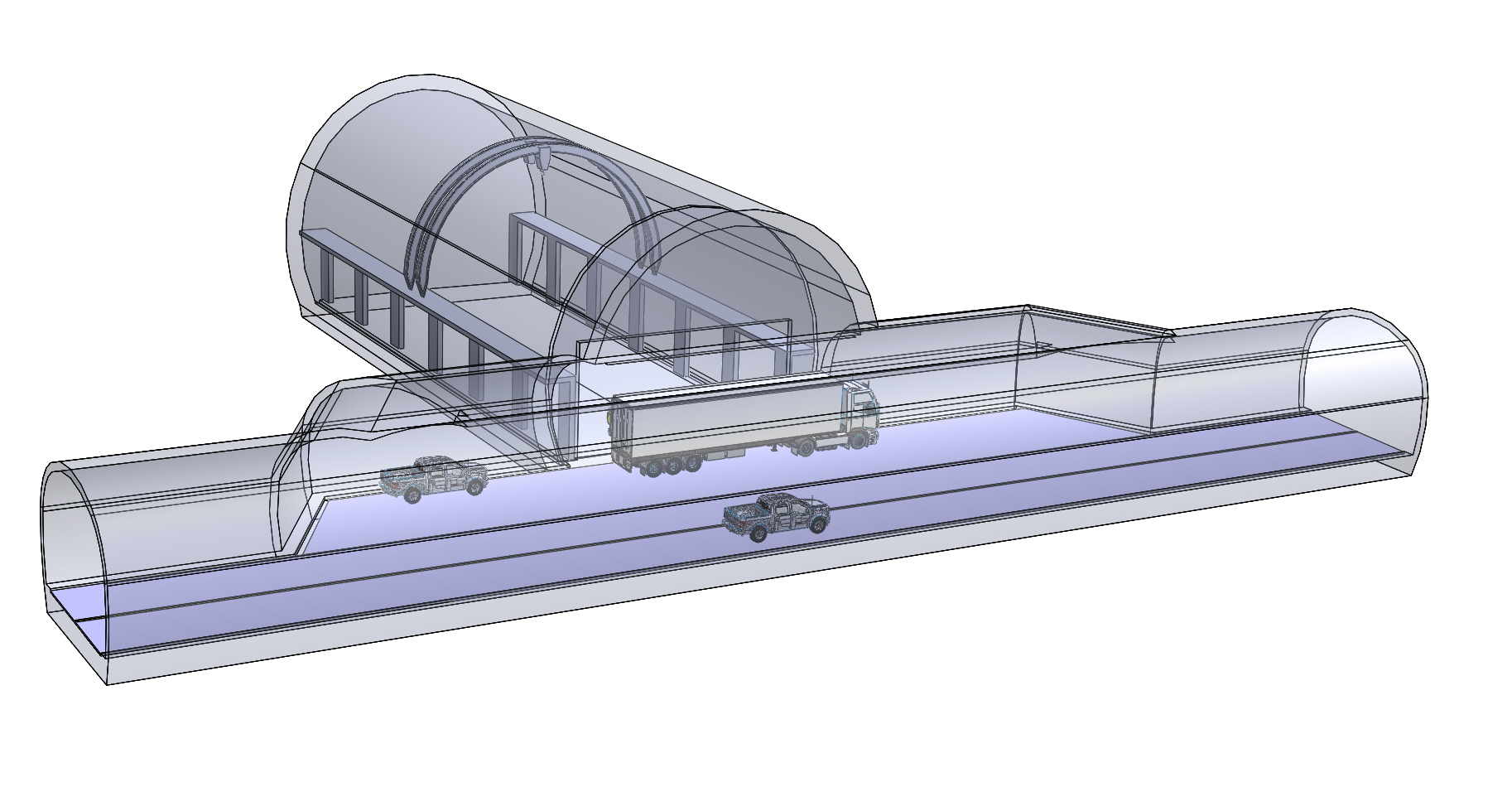}
\caption{\label{fig:infra} \small{Mock-up of a possible 600~m$^2$ laboratory (40x16x16 m$^3$) in the Huguenot tunnel. Courtesy: Joaqu\'in Venturino (CNEA), April 2023.}}
\end{figure}

For a small or medium size lab, the LSM concept~\cite{lsm}, where one has a deceleration line which stops on the side, is an appropriate design. The fact that one access tunnel to the lab is large allows one to stop a large truck in the proper space and then discharge the truck contents with a crane to move it inside the cavern. See Figure~\ref{fig:infra} for a minimal design of a possible 600~m$^2$ laboratory (40x16x16~m$^3$), based on the medium cavern design of ANDES~\cite{andessite}.

The future underground laboratory is currently being designed; It directly involves the company operating the Huguenot tunnel since earthworks and infrastructure construction are planned over the next five to ten years.

Apart from the operating surface and the digging work (in the tunnel) which involves large-scale building work, minimum services are necessary for the viability of the scientific installation. They are for example:
\begin{itemize}
\item Ventilation equipment capable of renewing the total volume of air in the laboratory every hour and of filtering the radon present in the air;
\item Air conditioning to maintain the laboratory at 21° Celsius. It is also necessary to control the relative humidity to guarantee the operation of equipment and the hygrothermal comfort of people;
\item Power supply equipment and generator;
\item Storage tanks and treatment plant for water and other effluents;
\item Provision of basic services to the experiments: Compressed air, water, data storage and processing, communications;
\item The computer center must have a fiber optic connection;
\item The medical care kit in the event of an accident;
\item Fire fighting systems, surveillance and security systems.
\end{itemize}

All these services require collaboration with industry, locally and regionally. Some state-of-the-art equipment are very typical of the discipline and can, sometimes, only be made with the help of international technological companies. 

%% file: conclusion.tex
\section{Conclusion}
PAUL is foreseen as an open international laboratory, a unique opportunity for Africa devoted to the development of a competitive science in the region. It has the advantage that the location, the Huguenot tunnel, exists already and the geology and the environment of the site is appropriate for an experimental facility.

PAUL will host international collaborations in experimental areas of great impact in physics, cosmology, geology, biology, material sciences, climate science etc. The activities of the laboratory will be accompanied by supporting centers and universities.

PAUL has a strong synergy with the radio astronomy probe of dark matter by South Africa's MeerKAT, HERA and SKA mid-array, through the WIMP annihilation and decay effects which eventually cascade down into synchrotron radiation. The strong synergy between ``ground'' (PAUL) and ``sky'' (SKA) can be combined to explore much larger parameter space with high-sensitivity, which may reveal some new physics beyond Standard Model.

The proposal will offer the chance to integrate with other facilities, laboratories and experiments and increase the potential of the region by increasing its academic activities, the formation of new human resources and the development of new basic research and technologies.

The initiative will be managed by a steering committee composed by representatives from South African universities and the South African Research Council, alongside experts from international facilities. The future laboratory will have a structure similar to an international facility such as Laboratoire Souterrain de Modane in France\cite{lsm} or Laboratori Nazionali Del Gran Sasso in Italy\cite{lngs}.

This is a unique opportunity to build an excellent deep underground laboratory, the only one in Africa, with a strong impact on regional integration.